\documentclass[journal]{IEEEtran}

\usepackage{verbatim}
\usepackage{amsfonts}
\usepackage{amssymb}
\usepackage{stfloats}
\usepackage{cite}
\usepackage{graphicx}
\usepackage{psfrag}
\usepackage{amsmath}
\usepackage{array}
\usepackage{epstopdf}
\usepackage{authblk}
\usepackage{graphicx} 
\usepackage{amsthm} 
\usepackage{lipsum}
\usepackage{verbatim} 
\usepackage{authblk}
\usepackage{mathtools}
\usepackage{cuted}
\usepackage{booktabs}

\usepackage{amsmath}
\usepackage{mathrsfs}

\usepackage{algorithmic}

\usepackage{array}

\ifCLASSOPTIONcompsoc
\usepackage[caption=false,font=normalsize,labelfont=sf,textfont=sf]{subfig}
\else
\usepackage[caption=false,font=footnotesize]{subfig}
\fi

\usepackage{url}
\usepackage{graphicx,amsmath,amssymb,amsfonts}
\usepackage{algorithmic,algorithm}

\usepackage[bookmarks=true,
colorlinks  =true,
breaklinks =true,
citecolor   =black,
linkcolor   =black,
urlcolor     =black,
]{hyperref}

\usepackage{setspace}

\makeatletter

\newcommand{\Rmnum}[1]{\expandafter\@slowromancap\romannumeral #1@}
\makeatother

\usepackage{framed} 

\usepackage{cancel}

\usepackage{booktabs}
\usepackage{tabularx,makecell,multirow}

\usepackage[table]{xcolor}
\usepackage{pifont} 

\usepackage{diagbox}

\begin{document}
	\bstctlcite{ref:BSTcontrol}
	
	\title{\fontsize{23.5 pt}{\baselineskip}\selectfont Joint Communication and Over-the-Air Computation for Semi-Federated Learning Towards Scalable AI\\ in Computing-Heterogeneous IoT Systems}
	
	\author{Wanli~Ni, and Hui~Tian
		\vspace{-3 mm}
		\thanks{Wanli Ni is with the Department of Electronic Engineering, Tsinghua University, Beijing 100084, China (e-mail: niwanli@tsinghua.edu.cn).}
		\thanks{Hui Tian is with the State Key Laboratory of Networking and Switching Technology, Beijing University of Posts and Telecommunications, Beijing 100876, China (e-mail: tianhui@bupt.edu.cn).}
}
	
	\maketitle
	
	\begin{abstract}
		The proliferation of Internet of Things (IoT) systems demands scalable artificial intelligence (AI) solutions that can operate in computing-heterogeneous environments with diverse hardware capabilities and non-independent and identically distributed data.
		This paper proposes a semi-federated learning ({\tt SemiFL}) framework that integrates centralized learning (CL) and federated learning (FL) to enable efficient model training across heterogeneous IoT devices. In {\tt SemiFL}, only devices with sufficient computational resources are designated for local model training (referred to as FL users), while the remaining devices transmit raw data to a base station (BS) for remote computation (referred to as CL users). This collaborative computing framework enables all IoT devices to participate in global model training, regardless of their computational capabilities and data distributions.
		Furthermore, to alleviate radio resource scarcity of {\tt SemiFL}, we propose a joint communication and over-the-air computation design that unifies wireless transmission and model aggregation.
		This approach reduces latency and enhances spectrum efficiency by allowing simultaneous communication and computation over the air.
		Furthermore, we design a transceiver architecture that integrates receive beamforming and successive interference cancellation to mitigate multi-user interference, ensuring reliable aggregation at the BS.
		Subsequently, two scenarios are examined to assess the efficacy of {\tt SemiFL} within IoT systems. In the first case, a reconfigurable intelligent surface is incorporated into the {\tt SemiFL} framework to dynamically adjust channel conditions, thereby enhancing communication reliability and mitigating aggregation distortion.
		In the second case, wireless power transfer is integrated into the {\tt SemiFL} system to facilitate the operation of battery-free IoT devices.
		Simulation results demonstrate that the proposed next-generation multiple access (NGMA)-based {\tt SemiFL} outperforms the fixed beamforming-based FL by 8\% and the AirComp-based FL by 6.4\%, demonstrating its superior learning efficiency and convergence performance.
	\end{abstract}
	

	\begin{figure*} [t]
		\centering
		\includegraphics[width=6 in]{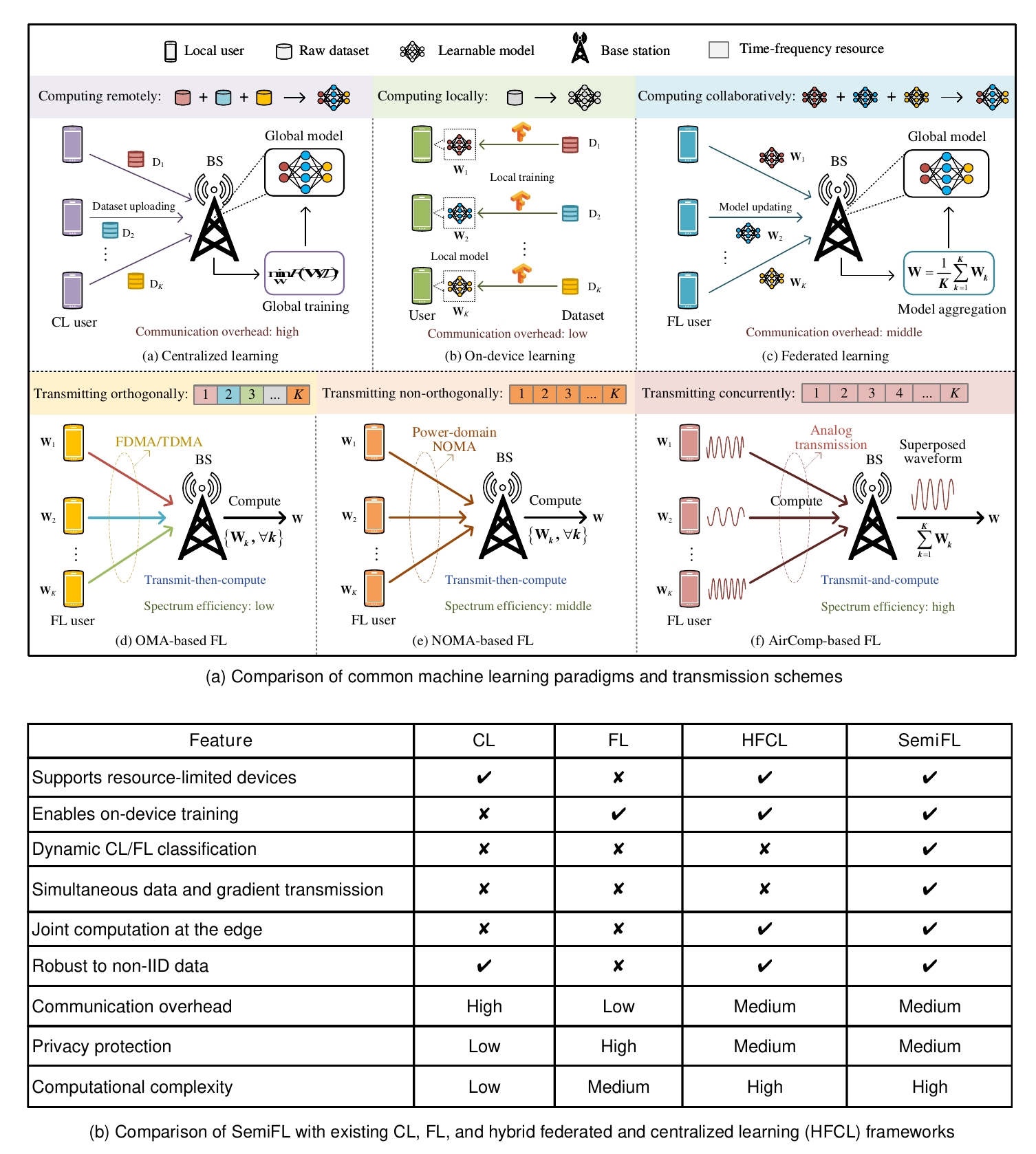}
		\caption{The first row illustrates three typical ML paradigms: CL, on-device learning, and FL. The second row demonstrates three common uplink transmission schemes for wireless FL, including OMA, NOMA, and AirComp. Specifically, CL is often deployed in the cloud-based system that has rich datasets and powerful computing capabilities. When implemented in IoT networks, CL entails key problems, such as high communication overhead, privacy risk, and inference delay. On-device learning allows all raw data to be processed on local devices. However, implementing on-device learning among low-end IoT devices faces issues such as insufficient computing capability, limited memory capacity, constrained energy supply. Although FL has presented huge potential in various intelligent IoT applications, it also faces challenges, such as unbalanced data distribution and computing-heterogeneous IoT devices.}
		\label{Fig1}
	\end{figure*}
	
	\vspace{-2 mm}
	\section{Introduction}
	The convergence of artificial intelligence (AI) and the Internet of Things (IoT) is driving a paradigm shift towards pervasive intelligence, enabling transformative applications such as Industry 4.0, autonomous transportation, and immersive metaverse services \cite{Xu2024Unleashing}. Modern IoT networks, comprising heterogeneous devices ranging from industrial sensors and surveillance cameras to wearable gadgets, generate vast amounts of multi-modal data that hold significant potential for real-time decision-making and predictive analytics \cite{Saad2025AGI}.
	However, achieving scalable AI at the wireless edge presents several formidable challenges: 1) managing computational heterogeneity among resource-constrained IoT devices, 2) ensuring communication efficiency in collaborative learning scenarios, and 3) mitigating performance degradation due to non-independent and identically distributed (non-IID) data.
	In the non-IID case, statistical heterogeneity across clients often leads to gradient divergence and class imbalance, which degrade convergence speed and model accuracy.

	Typical machine learning paradigms can be broadly categorized into three types, as depicted in Fig. \ref{Fig1}.
	Usually, centralized learning (CL), which aggregates raw data at a base station (BS) for cloud-based training, suffers from prohibitive communication overhead and privacy vulnerabilities \cite{Rahman2020Centralized}.
	On-device learning offers a solution by keeping data locally but is impractical for low-end IoT devices with limited computational power and energy resources \cite{Zhou2021On}.
	Federated learning (FL) partially overcomes these obstacles by facilitating distributed model training through periodic gradient aggregation \cite{Yonina2022SPM}.
	Yet, FL faces its own set of challenges, including stringent device requirements and performance degradation when handling non-IID data, which can significantly slow down convergence and reduce model accuracy \cite{Zhao2022FL, Ahmet2021Hybrid}.
	These limitations highlight the necessity for a hybrid framework that not only balances centralized and distributed computation but also optimizes wireless resource utilization.
	Namely, we need to design a promising solution to enable universal participation through dynamic role assignment based on computational capability, improve communication efficiency via joint communication and computation strategies, and enhance learning performance even under non-IID data conditions, thus achieving scalable and inclusive edge intelligence in diverse IoT environments.

	In this article, we present semi-federated learning ({\tt SemiFL}), an innovative framework that explores the integration of centralized and federated learning paradigms to address challenges in heterogeneous edge environments.
	Within the {\tt SemiFL} paradigm, FL users (devices with sufficient computing resources) perform local model training and transmit gradients to the BS. Conversely, CL users (resource-constrained devices) offload raw data to the BS for centralized computation.
	This two-tier architecture ensures full participation of heterogeneous devices while optimizing resource allocation. To address the communication bottlenecks in {\tt SemiFL}, we propose a joint communication and over-the-air computation design that unifies wireless transmission and model aggregation.
	The key contributions of this article are summarized as follows:
	\begin{itemize}
		\item
		We propose a novel {\tt SemiFL} framework that harmonizes CL and FL paradigms to address computing heterogeneity in IoT systems.
		By dynamically designating devices as FL users or CL users, {\tt SemiFL} ensures full participation of heterogeneous devices while optimizing resource utilization. To enable efficient data transmission, we develop a next-generation multiple access (NGMA) scheme that integrates non-orthogonal multiple access (NOMA) with over-the-air computation (AirComp). This scheme allows concurrent transmission of raw data (from CL users) and model gradients (from FL users) over shared radio resources, leveraging successive interference cancellation (SIC) and receive beamforming to mitigate multi-user interference, thus resulting in spectral efficiency improvements compared to conventional orthogonal multiple access (OMA) methods.
		\item
		We study the adaptability of {\tt SemiFL} within two complex IoT environments. Initially, a simultaneously transmitting and reflecting reconfigurable intelligent surface (STAR-RIS)-aided {\tt SemiFL} framework is presented, wherein the STAR-RIS dynamically modulates wireless channel conditions, effectively reducing aggregation distortion in non-line-of-sight scenarios, thereby markedly enhancing communication reliability for collaborative learning. Subsequently, a simultaneous wireless information and power transfer (SWIPT)-enabled {\tt SemiFL} framework is examined to accommodate battery-free IoT devices by integrating energy harvesting mechanisms. Collectively, these case studies illustrate {\tt SemiFL}’s compatibility with advanced wireless technologies and its resilience in practical deployment contexts.
		\item
		Simulation results confirm that the proposed NGMA-based {\tt SemiFL} achieves a test accuracy of 99.5\% after 100 communication rounds, outperforming the fixed beamforming-based FL by 8\% and the AirComp-based FL by 6.4\%. Furthermore, NGMA-based {\tt SemiFL} converges faster than both baselines, achieving over 95\% accuracy within 20 rounds, whereas fixed beamforming and AirComp-based FL require more than 40 rounds to reach similar levels.
		These gains are attributed to the hybrid architecture’s ability to balance local and remote computation, coupled with the NGMA scheme’s efficient resource utilization.
	\end{itemize}

	\section{SemiFL Framework and Transmission Design}
	Although FL has been regarded as a communication-efficient paradigm for edge intelligence in IoT networks, its performance remains vulnerable to two critical challenges: non-IID data across devices and computational heterogeneity arising from diverse hardware capabilities. These factors often lead to model divergence and straggler issues, particularly in resource-constrained IoT networks.
	In this section, we propose {\tt SemiFL}, a hybrid framework that seamlessly integrates CL and FL through a two-tier computing architecture.
	
	\begin{figure} [t]
			\centering
			\subfloat[Illustration of a generalized {\tt SemiFL} framework. Some users upload raw data to the BS for model computation, while others update model gradients to the BS for global aggregation.]{\label{Fig2a}
					\begin{minipage}[t]{0.45 \textwidth}
							\centering
							\includegraphics[width= 3.2 in]{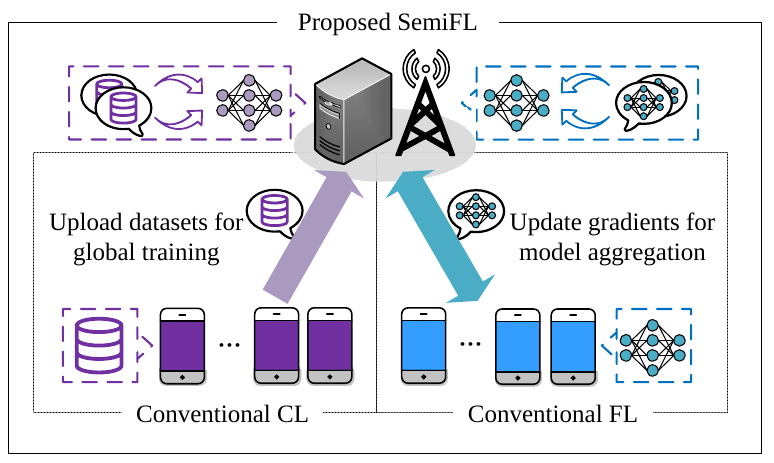}
						\end{minipage}
				} \\  
			\subfloat[The model training workflow of the proposed {\tt SemiFL} framework. The edge platform consists of a BS and a parameter server. The black dashed lines (blue solid lines) denote downlink (uplink) communications between the edge platform and IoT devices. The purple block indicates the data collection process. The blue (grey) blocks represent the local (global) computation processes.]{\label{Fig2b}
					\begin{minipage}[t]{0.45 \textwidth}
							\centering
							\includegraphics[width= 3.2 in]{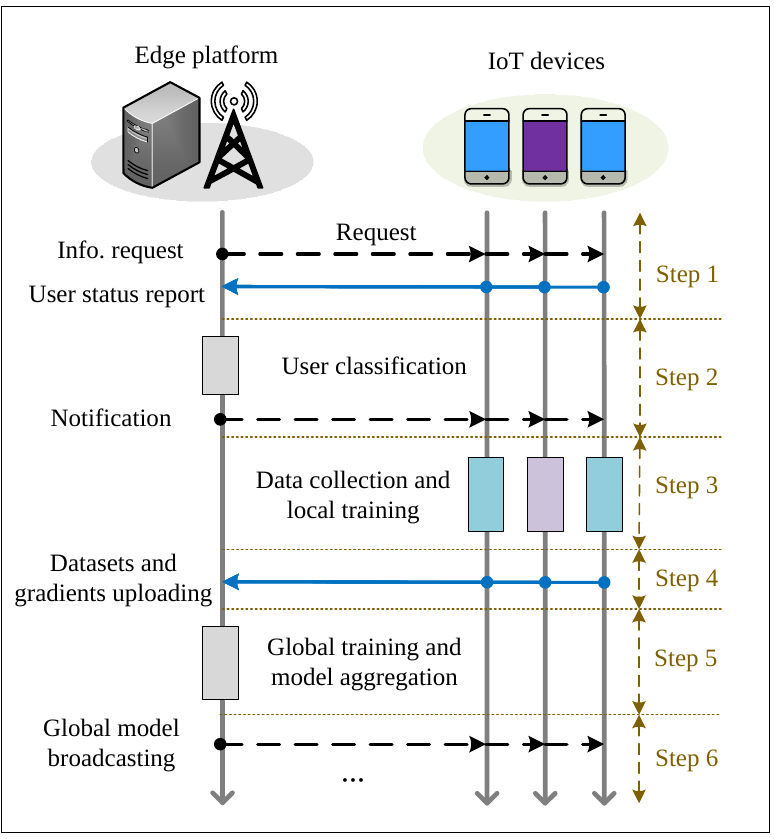}
						\end{minipage}
				}
			\caption{The proposed {\tt SemiFL} framework and its learning process.}
			\label{Fig2}
		\end{figure}
	
	\subsection{SemiFL Framework and Workflow} \label{section_SemiFL_procedure}
	\textbf{SemiFL framework:}
	In contrast to traditional approaches that exclude resource-constrained devices from model training, {\tt SemiFL} introduces a universal participation paradigm where all IoT nodes contribute to model optimization through a dual-mode collaborative framework.
	As shown in Fig.~\ref{Fig2a}, the framework partitions IoT devices into two groups: CL users and FL users.
	Specifically, devices with limited computational capabilities are classified as CL users and offload their raw data to an edge platform comprising a base station (BS) and an edge server. In contrast, the other devices are designated as FL users, who retain their local data and transmit only the model gradients \cite{Ahmet2021Hybrid}.
	Note that even if the data distributions of these CL users are non-IID and unbalanced, the edge server can approximately reconstruct a pseudo-IID dataset by aggregating data from multiple CL users, thereby improving global model generalization \cite{Han2024Semi}.
	In each round, the edge server (i) trains a CL model using the pooled CL data, (ii) aggregates the incoming FL gradients, and (iii) fuses the two into a single global model that is broadcast back to all device.
	This dual-layer design exploits both the raw-data bandwidth of CL users and the on-device computation of FL users, thereby optimizing the utilization of edge computing resources \cite{Zheng2023Semi}.
	Although prior works (e.g., \cite{Ahmet2021Hybrid} and \cite{Yoshida2020HybridFL}) have introduced similar hybrid learning methods, they either suffer from sequential execution bottlenecks, inflexible device roles, or lack of communication-efficient aggregation mechanisms. In contrast, {\tt SemiFL} introduces a collaborative framework that enables simultaneous data and gradient transmission, dynamic user classification based on computational capability, and joint NOMA and AirComp transmission.

	\begin{figure*} [t]
			\centering
			\includegraphics[width=6.5 in]{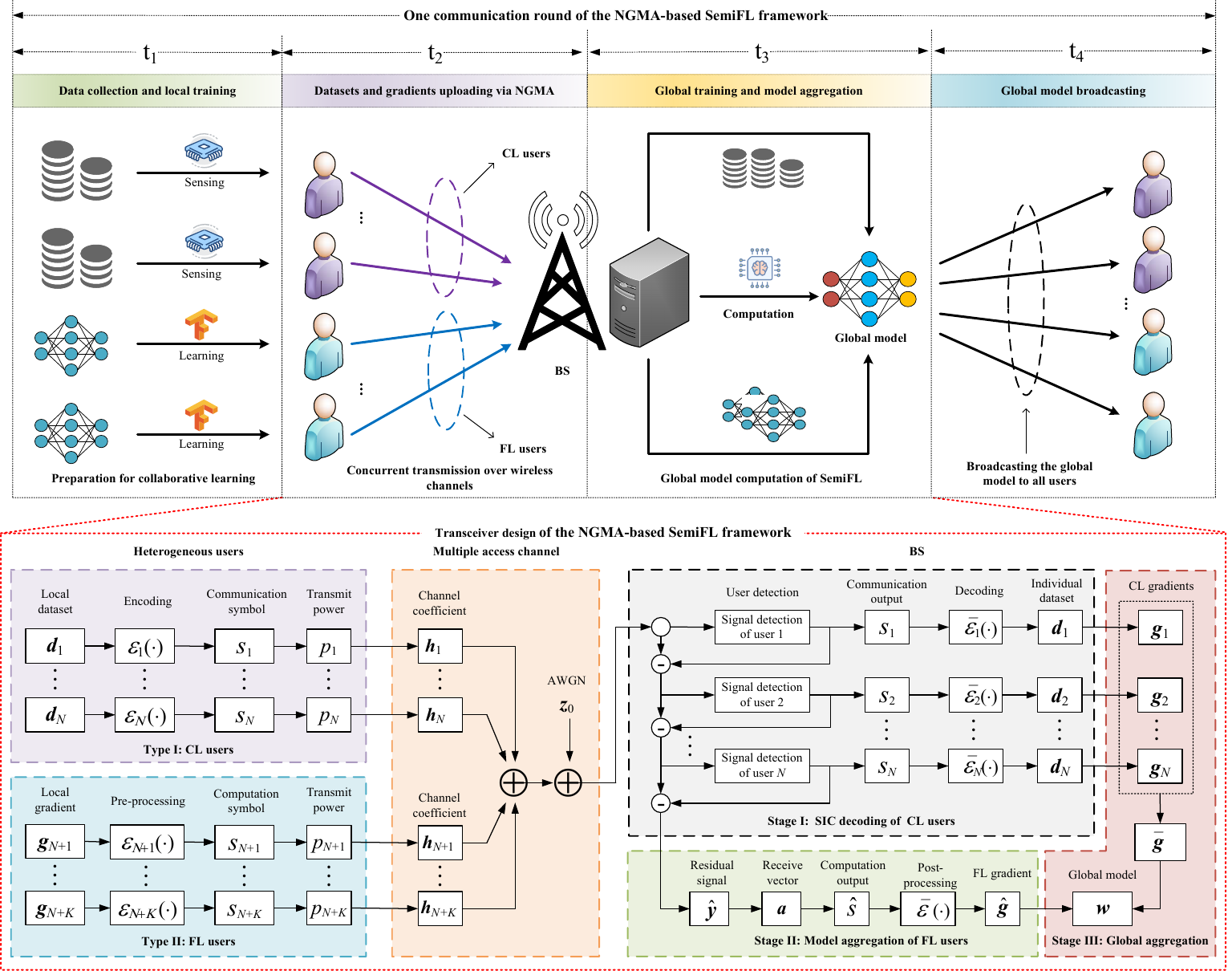}
			\caption{Schematic diagram of the proposed NGMA-based {\tt SemiFL} framework and its transceiver design. The top half illustrates the key steps of an NGMA-based {\tt SemiFL} framework in one communication round (i.e., Steps 3-6). The bottom half provides the corresponding transceiver design and the procedure of signal processing.}
			\label{Fig3}
		\end{figure*}

	\textbf{SemiFL workflow:}
	Fig.~\ref{Fig2b} summarizes the six-step {\tt SemiFL} pipeline executed in every communication round.
	\textit{Step 1 (Status polling):}
	The BS queries all devices to obtain their current communication, computation, and storage statuses. 
	\textit{Step 2 (Role assignment):}
	According to their computational capabilities, each device is tagged as either a communication-centric CL user (purple) or a computation-centric FL user (blue).
	These role assignments are then disseminated to all devices. 	
	\textit{Step 3 (Local activities):}
	Devices designated as FL users engage in on-device training and compute gradient updates, whereas CL users focus on acquiring new data samples. 
	\textit{Step 4 (Upload):}
	CL users transmit their raw datasets to the BS, while FL users upload their locally computed model updates. 	
	\textit{Step 5 (Model synthesis):}
	The BS trains a CL model on the collected data, merges it with the aggregated FL model, and produces a new global model~\cite{Yoshida2020HybridFL}.
	\textit{Step 6 (Distribution):}
	The newly synthesized global model is distributed to all devices for subsequent rounds.
	This iterative process continues until convergence or another stopping criterion is met, e.g., the validation loss stabilizes within $\pm 0.001$ for three consecutive rounds or the maximum epoch limit (e.g., 100 rounds) is reached.
	
	\textbf{Computational complexity:}
	In CL, the server performs all training, leading to high computational load at the edge server: $\mathcal{O}(Nd)$, where $N$ is the number of samples and $d$ is the model dimension.
	In FL, each active device computes local updates: $\mathcal{O}(K d n_k)$, where $K$ is the number of participating devices and $n_k$ is the local dataset size.
	In SemiFL, both FL users (performing local training) and CL users (transmitting raw data) contribute: $\mathcal{O}(K d n_k) + \mathcal{O}(M d \bar{n})$, where $M$ is the number of CL users and $\bar{n}$ is the average data size per user.

	\textbf{Convergence analysis:}
	Based on the convergence analysis presented in our previous work \cite{Han2024Semi}, we can know that the proposed {\tt SemiFL} converges to a stationary point with a rate of $\mathcal{O}(1/{T})$ under standard assumptions of smoothness, bounded variance, and bounded gradient dissimilarity.
	This indicates that {\tt SemiFL} achieves a linear convergence rate comparable to conventional FL, despite the hybrid nature of data and model updates from heterogeneous devices. The analysis also explicitly confirms that the proposed hybrid learning paradigm remains stable and effective in computing-heterogeneous IoT networks.
	
	\subsection{Multiple Access and Transceiver Design for SemiFL}
	\label{section_NGMA}
	Traditional communicate-then-compute approaches are misaligned with {\tt SemiFL}’s learning objectives, thereby necessitating a paradigm shift from communication-centric to learning-centric transmission strategies. This shift requires leveraging the inherent interplay between communication and learning within {\tt SemiFL} to enhance resource efficiency.
	In this subsection, we introduce a NGMA scheme specifically designed to meet the uplink transmission demands of {\tt SemiFL}, accompanied by a comprehensive description of its transceiver architecture. Within the uplink transmission of {\tt SemiFL}, two distinct paradigms exist for the uploading of datasets by CL users and gradients by FL users.
	The first paradigm, termed \textit{separate design}, involves the allocation of bandwidth or temporal resources to facilitate independent transmission of data and gradients \cite{Ahmet2021Hybrid}. However, this approach results in suboptimal spectral efficiency, as CL data transmissions typically demand considerable bandwidth.
	The second paradigm, referred to as \textit{unified design}, employs a joint framework that concurrently serves both user categories, thereby addressing the conflicting objectives of maximizing data rates for CL users and minimizing computation distortion for FL users \cite{Ni2022Integrating}.
	Existing approaches inadequately fail to reconcile these heterogenous objectives, thereby motivating our NGMA scheme, which synergistically integrates NOMA and AirComp techniques to realize a spectrum-efficient joint communication and computation scheme to improve the uplink transmission efficacy of {\tt SemiFL}.
	
	As illustrated in Fig.~\ref{Fig3}, the proposed NGMA scheme achieves a seamless integration of heterogeneous learning paradigms by unifying NOMA-based raw data transmission for CL users and AirComp-based gradient aggregation for FL users within a shared wireless spectrum.
	Exploiting the superposition property of over-the-air channels, NGMA enables simultaneous uplink transmissions from both CL and FL devices, thereby maximizing spectral efficiency and supporting scalable edge intelligence.
	To resolve interference from concurrent transmissions, the BS employs SIC in a channel-gain-ordered manner: users with stronger channel conditions are decoded first, their signals are reconstructed and subtracted from the composite received signal, allowing weaker users to be successively recovered. This hierarchical decoding aligns perfectly with the dual objectives of {\tt SemiFL}:
	\begin{itemize}
		\item 
		For CL users, individual signal decoding ensures accurate reconstruction of raw data for centralized model training.		
		\item 
		For FL users, signal superposition naturally performs analog gradient aggregation via AirComp, preserving the averaged model update without explicit separation.
	\end{itemize}
	At the transmitter side, CL users encode their local datasets into communication symbols, while FL users precode their gradients into computation symbols, both tailored for reliable over-the-air delivery.
	At the receiver, the BS first decodes and removes the CL user signals using SIC, then directly extracts the over-the-air aggregated gradient from the remaining signal. Finally, the global model is refined by jointly utilizing the reconstructed CL data and the aggregated FL gradients, realizing a truly unified framework for hybrid model updating.

	\begin{figure} [t]
		\centering
		\includegraphics[width=3.5 in]{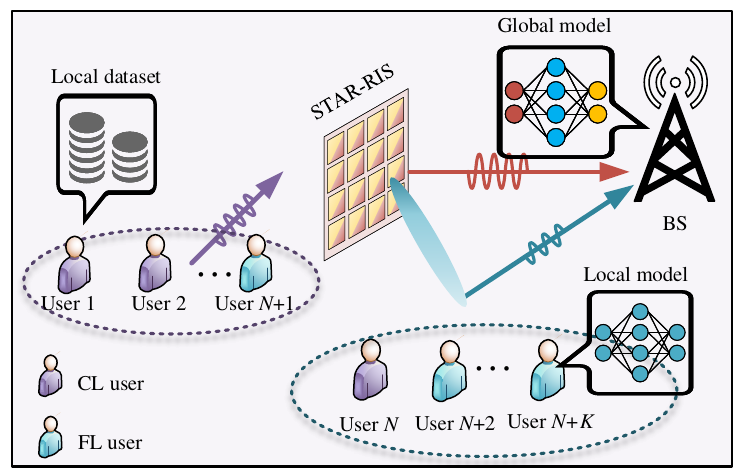}
		\caption{A case of {\tt SemiFL} in STAR-RIS-aided IoT systems. The STAR-RIS enhances data accessibility and manages interference, allowing for more reliable gradient updates and dataset uploads, which contribute to the overall learning performance of {\tt SemiFL}.}
		\label{Fig4}
	\end{figure}
	
	\section{Case Studies of SemiFL in IoT Systems}
	
	\subsection{Case 1: SemiFL in STAR-RIS-Aided IoT Systems}
	The proposed NGMA-based {\tt SemiFL} necessitates meticulous coordination of concurrent transmissions from both CL and FL users. This coordination is achieved by strategically arranging the decoding order based on the prevailing channel conditions. However, traditional wireless environments are inherently static, offering limited flexibility for on-demand modifications, and typically require complex transceiver designs for compensation.
	Recently, the advent of reconfigurable intelligent surfaces (RIS) has revolutionized this paradigm by treating the wireless environment as a dynamic and controllable optimization variable. By finely tuning the amplitude and phase shifts of incident signals, RIS technology can unlock substantial performance enhancements, including improved spectrum and energy efficiency. Unlike conventional RIS, which solely reflects signals, the STAR-RIS extends coverage by supporting both reflection and refraction, enabling full-dimensional signal propagation.
	
	To further elevate the learning performance of {\tt SemiFL} within heterogeneous IoT environments, a STAR-RIS-aided {\tt SemiFL} framework has been introduced \cite{Ni2022SemiFL}. This framework enhances communication efficiency for CL users and mitigates aggregation distortion for FL users by leveraging the STAR-RIS for effective interference management and channel alignment. As depicted in Fig. \ref{Fig4}, the STAR-RIS is strategically positioned near edge devices to facilitate signal relaying to the BS for efficient message decoding and model aggregation.
	Compared to the standalone {\tt SemiFL} framework discussed in Section \ref{section_NGMA}, where only transmission and reception schemes are optimized, the STAR-RIS-aided {\tt SemiFL} framework offers several distinct advantages. 
	Firstly, the STAR-RIS can be harnessed to constructively or destructively modify the channel quality experienced by individual users by adjusting reflection/refraction coefficients and optimizing its deployment location \cite{Ni2022STAR}. This flexibility allows for a more adaptive decoding order design for CL users employing NOMA for local dataset uploads.
	Secondly, the STAR-RIS introduces additional degrees of freedom that can be exploited to align the weighted sum of input parameters, thereby reducing aggregation distortion for FL users utilizing AirComp for local model updates. Furthermore, the STAR-RIS seamlessly integrates with existing systems, making it readily deployable in heterogeneous IoT networks without the need for additional time slots or new protocol designs.
	Consequently, the learning procedure and transceiver design of the STAR-RIS-aided {\tt SemiFL} framework remain largely consistent with those of the standalone version, with the added consideration of optimizing the STAR-RIS configuration. Experimental results presented in \cite{Ni2022SemiFL} and \cite{Ni2022STAR} demonstrate that the STAR-RIS can significantly enhance the spectral efficiency of {\tt SemiFL} while accelerating the convergence speed of model training.

	\begin{figure*} [t]
		\centering
		\includegraphics[width=6.7 in]{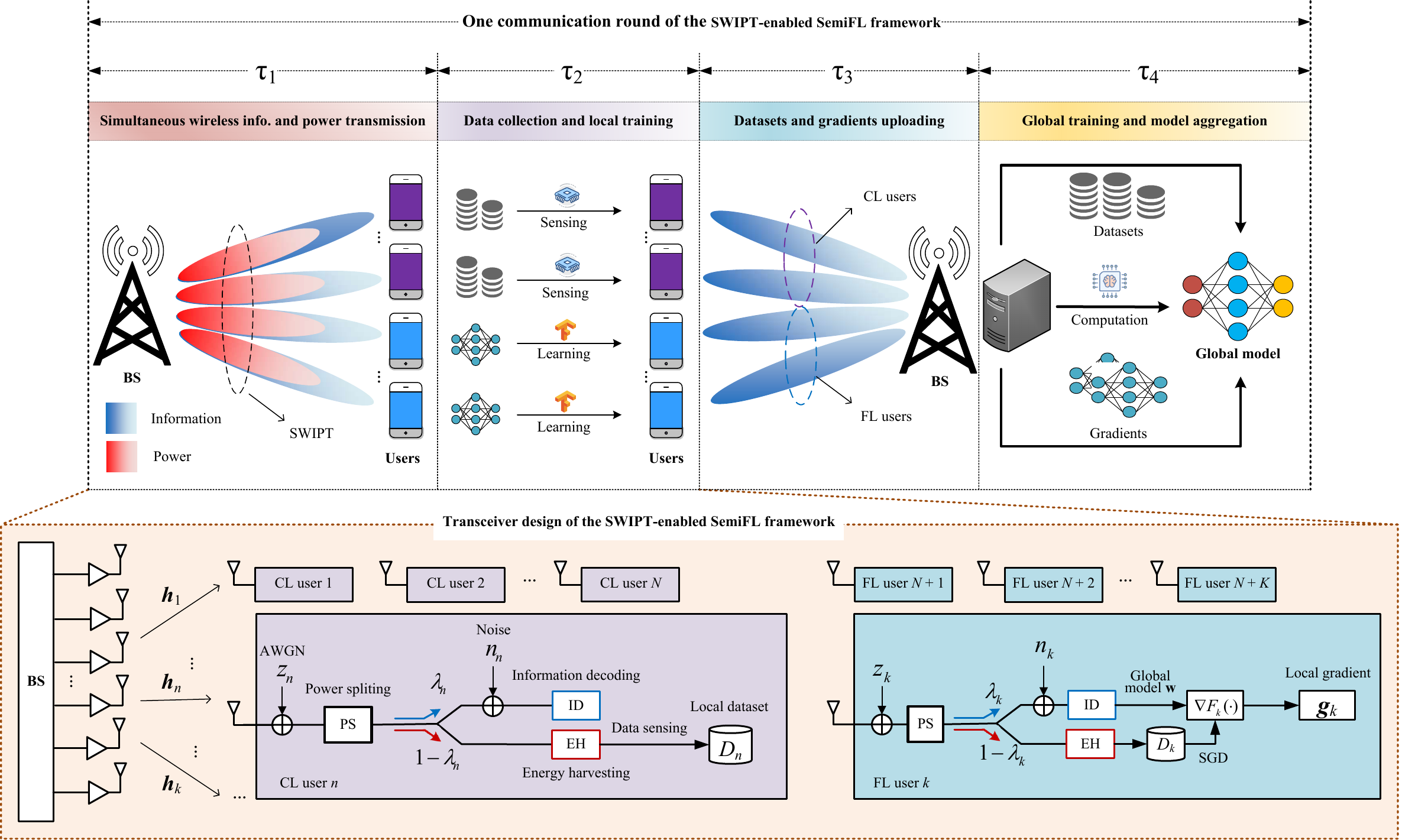}
		\caption{An illustration of the SWIPT-enabled {\tt SemiFL} framework and its transceiver design. These IoT devices harvest energy from the BS to power their operations, including data collection, local model training, and uplink transmission.}
		\label{Fig5}
	\end{figure*}
	
	
	\subsection{Case 2: SemiFL in SWIPT-Enabled IoT Systems}
	For IoT devices operating with limited energy capacity, the expenses associated with maintenance can be notably high.
	Thus, tackling the energy constraints of these devices is of utmost significance.
	A proven strategy to combat energy scarcity in distributed IoT networks is to endow IoT devices with the capability to harvest energy \cite{Zhang2024Deep}, enabling them to perpetually capture radio frequency energy emitted by the associated BS.
	The fusion of {\tt SemiFL} with energy harvesting technology empowers battery-less IoT devices with nearly perpetual learning capabilities, contingent upon the durability of their hardware components.
	Therefore, to bypass the limitations of energy storage in IoT devices, we present a SWIPT-enabled {\tt SemiFL} framework shown in Fig. \ref{Fig5}.
	This framework ensures a stable and controllable transfer of wireless energy from the BS to devices with constrained battery capacity.
	
	The core process of SWIPT-enabled {\tt SemiFL} follows the harvest-then-learn protocol.
	Specifically, during the initial time slot, wireless information (comprising global model parameters) and power are concurrently transmitted to all devices over a shared frequency band. 
	Subsequently, in the second time slot, after harvesting sufficient energy from the BS, these IoT devices commence data sample collection and initiate local learning. In the third time slot, the newly acquired data from CL users and the updated model parameters from FL users are relayed to the BS.
	Finally, in the fourth time slot, the decoded CL datasets are leveraged to utilized the prediction accuracy of the aggregated FL model.	
	It is noteworthy that the procedures of the last three time slots correspond to Steps 3–5 outlined in Section \ref{section_SemiFL_procedure}, and hence, a detailed description of the uplink transmission is omitted here for brevity.
	In accordance with the harvest-then-learn protocol, IoT devices, upon harvesting energy from the BS, proceed to collect data samples, perform local learning, and transmit both the sensed data and updated model parameters to the BS.
	While energy harvesting confers benefits such as self-sustainability and operational flexibility, the development of a high-performance SWIPT-enabled {\tt SemiFL} system introduces novel challenges. For instance, the hardware architecture of IoT devices must be adapted to integrate information decoding and energy harvesting functionalities seamlessly. Furthermore, optimizing the power splitting ratio, denoted by \(\rho\), at the IoT devices is critical to balancing the trade-off between information transmission and power transfer effectively.
	
\begin{figure*} [t!]
	\centering
	\subfloat[Training loss vs. the number of communication rounds]{\label{Fig6a}
		\begin{minipage}[t]{0.45 \textwidth}
			\centering
			\includegraphics[width= 3.3 in]{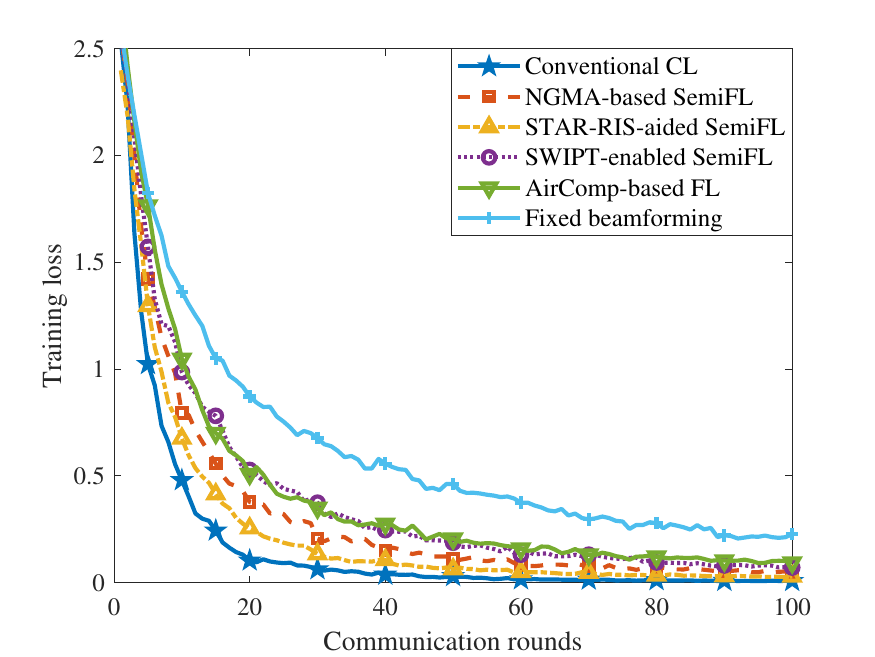}
		\end{minipage}
	}  
	\subfloat[Test accuracy vs. the number of communication rounds]{\label{Fig6b}
		\begin{minipage}[t]{0.45 \textwidth}
			\centering
			\includegraphics[width= 3.3 in]{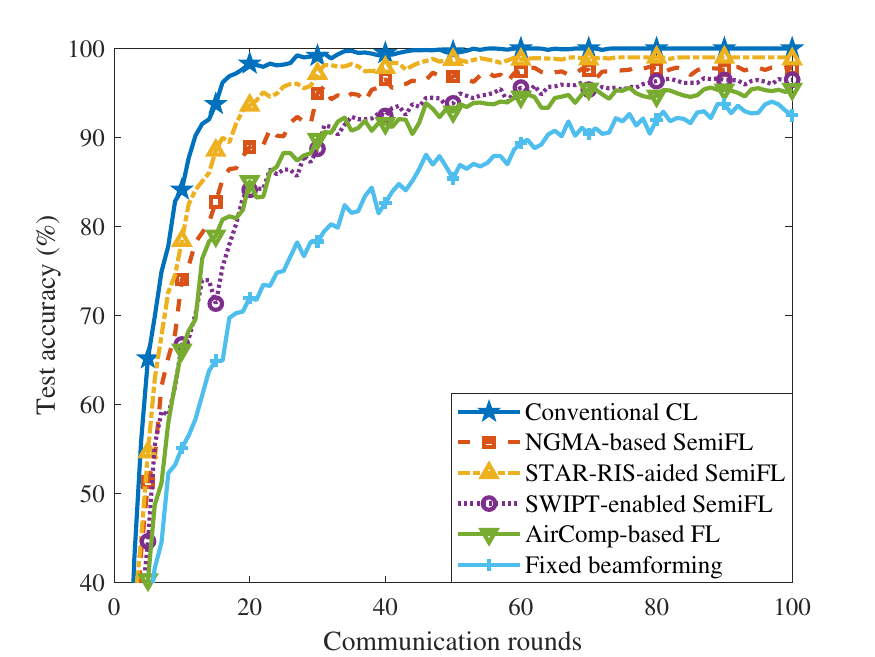}
		\end{minipage}
	}
	\caption{Learning performance of training a CNN on the MNIST dataset: (a) training loss, and (b) test accuracy.}
	\label{Fig6}
\end{figure*}	
	
	\subsection{Simulation Results}
	To evaluate the effectiveness of our proposed {\tt SemiFL} frameworks, we consider an IoT network comprising two CL users and eight FL users.
	Following similar setups with our previous works \cite{Ni2022STAR, Ni2022SemiFL}, we employ a six-layered convolutional neural network (CNN) as the classifier model. The CNN model is tasked with performing handwritten-digit recognition on the MNIST dataset.
	We adopt a Rayleigh fading channel model to simulate the wireless propagation environment, with path loss exponent set to $3.5$ and additive white Gaussian noise (AWGN) with a variance of $\delta^2 = -80$ dBm.
	The energy harvesting efficiency at each IoT device is set to $\eta = 0.6$.
	The number of communication rounds is set to $100$, with each round consisting of local model training (for FL users), data uploading (for CL users), and global model aggregation at the BS. The learning rate is initialized to $0.01$ and decayed by a factor of $0.95$ every $10$ rounds.
	We compared our {\tt SemiFL} schemes against three benchmarks:
	1) Conventional CL: All users transmit their entire datasets to the BS for centralized training.
	2) AirComp-based FL: All users perform local model training and upload their model gradients using AirComp;
	3) Fixed beamforming: The transceiver design remains static across communication rounds, without dynamic adaptation to channel conditions.
	
	Fig. \ref{Fig6} illustrates the training loss and test accuracy achieved by the different schemes over 100 communication rounds. The proposed STAR-RIS-aided {\tt SemiFL} framework demonstrates remarkable performance, closely approaching that of the conventional CL scheme, which enjoys ideal model updates with centralized processing. This observation underscores the efficacy of integrating raw data from computing-limited CL users into the global model training process, thereby enhancing the overall learning performance.
	Furthermore, our {\tt SemiFL} frameworks significantly outperform the AirComp-based FL and fixed beamforming schemes. This superiority can be attributed to two key factors. First, the inclusion of raw data from CL users enriches the global dataset, mitigating the impact of non-IID data distributions among FL users. Second, the dynamic adjustment of the transceiver design, facilitated by the NGMA scheme, effectively alleviates the negative effects of noisy fading channels on learning performance.
	Notably, the STAR-RIS-aided {\tt SemiFL} framework achieves the lowest training loss and highest test accuracy among all considered schemes. This improvement is primarily due to the STAR-RIS's ability to enhance the signal quality of both CL and FL users by intelligently reflecting and refracting incident signals, thereby improving the overall throughput of NOMA-based CL users and reducing the aggregation error of AirComp-based FL users.
	The scalability and robustness of {\tt SemiFL} under varying network conditions and system configurations have been thoroughly investigated in our prior studies \cite{Han2024Semi, Zheng2023Semi}, and are therefore omitted here for brevity.

	\section{Open Issues and Research Opportunities}
	\subsection{Theoretical Foundations and Performance Limits}
	Existing convergence analyses for {\tt SemiFL} often assume idealized digital communication links or additive Gaussian noise channels. However, practical wireless systems introduce non-linear distortions (e.g., power amplifier clipping), time-varying fading, and rank-deficient aggregation matrices due to spatially correlated devices. A rigorous theory is needed to quantify how fading statistics (e.g., Rayleigh, Rician), transceiver hardware impairments (e.g., I/Q imbalance), and partial device participation jointly degrade model prediction accuracy and generalization bounds.
	Moreover, differential privacy in {\tt SemiFL} relies on a delicate balance between channel-induced noise (natural privacy from AirComp) and artificial noise (added for algorithmic privacy). Characterizing the Pareto frontier between privacy budgets, convergence accuracy, and energy efficiency remains an open information-theoretic problem.
	
	\subsection{Asymmetric Channels and Asynchronous Update}
	In heterogeneous networks (e.g., satellite-terrestrial integration), IoT devices often encounter asymmetric channel capacities, posing significant challenges to the deployment of SemiFL. Currently, there is a lack of theoretical frameworks capable of accurately bounding generalization errors when model parameters are transmitted over channels characterized by unequal signal-to-noise ratios (SNRs) or varying latencies.
	This gap hinders the ability to predict and optimize model performance in real-world scenarios. Moreover, most existing {\tt SemiFL} protocols are predicated on synchronized updates among devices. However, the presence of stragglers, attributed to factors like channel fading or limited battery life, can significantly degrade overall system efficiency.
	Designing asynchronous protocols that allow devices to transmit local updates opportunistically, such as when channel conditions are favorable, without compromising the convergence properties of the global model, represents a critical research challenge. 
	
	\subsection{System-Level Design and Prototyping}
	The majority of existing studies on AirComp-based FL assume the use of linear transceivers and ideal synchronization conditions. However, real-world radio systems are subject to impairments such as phase noise, nonlinear distortions introduced by power amplifiers, and sampling rate mismatches, all of which contravene these idealized assumptions. Consequently, the development of an open radio access network (RAN) testbed employing software-defined radios (SDRs), such as the USRP N310, alongside FPGA-based prototypes, is imperative. Such a testbed would facilitate the empirical quantification of these hardware impairments and enable the validation of compensation techniques, including digital pre-distortion and carrier frequency offset correction.
	In addition to addressing hardware challenges, the design of cross-layer protocols is crucial. For instance, integrating {\tt SemiFL} with 5G/6G new radio (NR) numerologies (e.g., flexible subcarrier spacing) or Wi-Fi 7 multi-link operation could enable dynamic resource allocation.
	Open-source frameworks should be extended to support {\tt SemiFL}-specific primitives (e.g., AirComp layers, channel-aware gradient compression).

	\subsection{Regulatory for Over-the-Air Model Training}
	Current spectrum regulations assume human-centric traffic. Training AI models over-the-air produces persistent, wideband, non-orthogonal transmissions that challenge traditional interference models. Engaging with regulators to carve out experimental bands or define new etiquette protocols is an urgent, under-studied issue.
	Finally, standardization efforts are needed to define benchmarks for {\tt SemiFL} in real-world scenarios (e.g., vehicular networks, IoT). Collaborative initiatives between academia and industry (e.g., OpenAirInterface, O-RAN Alliance) could accelerate the transition from theory to deployable systems.
	
	\subsection{AI-Based Dynamic Client Role Assignment in SemiFL}
	To enhance the adaptability and scalability of {\tt SemiFL} in large-scale, heterogeneous IoT deployments, a critical research direction is the development of AI-based real-time client role assignment mechanisms that dynamically classify devices as FL participants, CL offloaders, or idle nodes based on their evolving computational resources, data quality, and energy constraints. Unlike static role assignments, this dynamic approach would allow SemiFL to optimize resource utilization while maintaining model performance across diverse IoT environments.
	A key challenge lies in designing lightweight, on-device AI agents capable of predicting optimal roles with minimal computational overhead. These agents must balance accuracy and efficiency to avoid draining constrained IoT devices while ensuring that role assignments promote global model convergence and fairness. 
	
	\subsection{SemiFL for Foundation Model Fine-Tuning in IoT Systems}
	The integration of foundation models (e.g., GPT, Llama, and Qwen) into IoT systems presents a transformative opportunity for edge intelligence \cite{Golec2025AI}. However, deploying and adapting these models in real-world IoT environments faces critical challenges, e.g., computational heterogeneity across devices, data heterogeneity due to diverse and non-IID data distributions, and prohibitive communication overhead during model updates.
	Our {\tt SemiFL} presents a two-tier fine-tuning paradigm for foundation models. Specifically, IoT devices with sufficient computational resources (FL users) compute gradients for specific layers or low-rank adapters, while resource-constrained devices (CL users) offload raw data for centralized processing, and then edge servers act as intermediate coordinators to perform parameter-efficient fine-tuning. Nevertheless, it is challenging to balance model personalization for device-specific tasks and generalization across heterogeneous IoT environments.
	
	\section{Conclusions}
	In this article, we have proposed a {\tt SemiFL} framework that effectively integrates the strengths of both CL and FL into a unified and harmonized framework. This innovative approach enables all IoT devices, regardless of their computational capabilities and data distributions, to collaboratively train a shared model.
	To further enhance the performance of {\tt SemiFL}, we have designed a learning-centric NGMA uplink transmission scheme. This scheme seamlessly integrates communication and computation over the air, enabling concurrent transmission of raw data and model parameters in a spectrum-efficient manner.
	Moreover, we have incorporated two emerging techniques, namely STAR-RIS and SWIPT, into the {\tt SemiFL} framework, makes it a promising candidate for resource-constrained IoT networks. 
	Experimental results showed that the {\tt SemiFL} framework, particularly when aided by STAR-RIS, outperforms AirComp-based FL and fixed beamforming approaches in terms of both training loss and test accuracy.
	
	\section*{Acknowledgement}
	The work of Wanli Ni was supported by the National Natural Science Foundation of China (Grant No. 62501351), the Postdoctoral Fellowship Program of the China Postdoctoral Science Foundation (Grant No. GZB20240386), and the China Postdoctoral Science Foundation (Grant No. 2024M761669).

	\bibliographystyle{IEEEtran}
	\bibliography{IEEEabrv,ref}

	\begin{IEEEbiographynophoto}{Wanli Ni} (Member, IEEE) (niwanli@tsinghua.edu.cn) is currently a Postdoctoral Researcher with the Department of Electronic Engineering, Tsinghua University, China.
	He received the B.Eng. and Ph.D. degrees from the Beijing University of Posts and Telecommunications, China, in 2018 and 2023, respectively.
	From 2022 to 2023, he was a Visiting Student at Nanyang Technological University, Singapore. His research interests include federated learning, semantic communications, large AI models, and reconfigurable intelligent surface.
	During his PhD studies, he received the National Scholarship twice and was recognized four times as an IEEE Exemplary Reviewer.
	In 2023, he received the Outstanding Doctoral Dissertation Award from the China Society of Electronics Education and was honored as an Outstanding Graduate of Beijing Municipality.
	In 2025, he was named to the World’s Top 2\% Scientists list.
	\end{IEEEbiographynophoto}
	
	\begin{IEEEbiographynophoto}{Hui Tian} (Senior Member, IEEE) (tianhui@bupt.edu.cn) received the M.S. and Ph.D. degrees from Beijing University of Posts and Telecommunications, China, in 1992 and 2003, respectively. Currently, she is a Professor with the School of Information and Communication Engineering, BUPT. Her research interests include radio resource management in 5G/6G networks, mobile edge computing, cooperative communication, mobile social networks, and Internet of Things.
	\end{IEEEbiographynophoto}
\end{document}